\documentclass[12pt]{iopart}
\usepackage{iopams}
\usepackage{graphicx}
\usepackage{bold-extra}
\usepackage{listings}
\usepackage{bm,color}
\usepackage[usenames,dvipsnames]{xcolor}
\usepackage{bbm}

\definecolor{darkgreen}{rgb}{.1,.4,.15}
\definecolor{darkred}{rgb}{.5,.1,.2}
\definecolor{darkblue}{rgb}{.2,.3,.6}

\lstdefinelanguage{modelica}{%
    alsoletter={...},%
    morekeywords=[1]{assert,extends,external,annotation,block,class,connector,constant,discrete,%
    else,elseif,elsewhen,end,equation,exit,final,flow,for,%
    function,if,in,inner,input,import,loop,model,nondiscrete,outer,%
    output,package,parameter,partial,record,redeclare,replaceable,return,%
    size,terminate,then,type,when,while,algorithm,equation,%
    protected,public,and,false,not,or,true},
    morekeywords=[2]{
        },%
    morekeywords=[3]{
        abs,acos,asin,atan,atan2,connect,cos,cosh,cross,der,edge,exp,%
        initial,noEvent,pre,reinit,sample,sign,sin,sinh,tan,tanh,terminal,%
        start,Real,Integer,Boolean,String, Complex},%
    comment=[l][\color{darkgreen}]{//}, 
    morecomment=[s][\color{darkgreen}]{/*}{*/}, 
    morestring=[b][\color{darkgreen}]{'}, %
    morestring=[b][\color{darkgreen}]{"},
}[keywords,comments,strings]

\begin{document}

\title{Transformation of quantum photonic circuit models by term rewriting}

\author{Gopal Sarma, Ryan Hamerly, Nikolas Tezak, Dmitri S.~Pavlichin, Hideo Mabuchi}
\address{Edward L.~Ginzton Laboratory, Stanford University, Stanford, CA 94305, USA}

\date{\today}

\begin{abstract}
The development of practical methods for synthesis and verification of complex photonic circuits presents a grand challenge for the nascent field of quantum engineering. Of course, classical electrical engineering provides essential foundations and serves to illustrate the degree of sophistication that can be achieved in automated circuit design. In this paper we explore the utility of term rewriting approaches to the transformation of quantum circuit models, specifically applying rewrite rules for both reduction/verification and robustness analysis of photonic circuits for autonomous quantum error correction. We outline a workflow for quantum photonic circuit analysis that leverages the Modelica framework for multi-domain physical modeling, which parallels a previously described approach based on VHSIC Hardware Description Language (VHDL).
\end{abstract}



\maketitle

\noindent A broad range of current research projects in photonics and quantum electronics are devoted to the development of transducers, logic gates, and related components based on quantum-mechanical device physics~\cite{NJPFocus}. In order to realize the long-term vision of advanced technology based on complex networks of interconnected quantum devices, equal attention will need to be paid to developing the quantum theory of autonomous (embedded) photonic, optomechanical and and optoelectronic circuits~\cite{Teza12}. This should be true not only for the most ambitious paradigm of true quantum computing and communication, but also for engineering approaches that seek to leverage coherent photonic, electronic or spintronic resources for quantitative improvements in the speed and/or energy efficiency of classical sensing~\cite{Thom11}, information processing~\cite{Mabu11b} and communication~\cite{Hall11}.

Within the realm of photonics, and by extension for closely related systems in circuit quantum electrodynamics (circuit QED) and quantum optomechanics, a formalism based on quantum stochastic differential equations (QSDEs)~\cite{HP84,Barc06} can be used to construct time-domain Heisenberg picture models for circuits comprising multiple quantum input-output components with cascade and feedback interconnections based on coherent signal propagation through optical (or microwave) waveguides. The electromagnetic signal propagating from a component output port, through a waveguide, to an input port makes it possible for the evolution of one component to influence that of another, or can modify the dynamics of a single component via coherent feedback. A particularly convenient modeling approach developed by Gough and James~\cite{GJ09a}, which generalizes earlier results of Carmichael~\cite{Carm93} and Gardiner~\cite{Gard93}, uses a parameter triple $(S,L,H)$ to represent the internal dynamics and input-output couplings of each component together with series and concatenation products to express circuit models in terms of the port connection topology. Here $S$ is a scattering matrix that describes direct couplings among the input and output ports of a component, $L$ is a vector of operators that describes the way that the internal variables of a component couple to input and output fields, and $H$ is the Hamiltonian operator describing the intrinsic dynamics of the internal variables. In general the matrix elements of $S$ can be either c-numbers or operators on the component's internal Hilbert space, and the $(S,L,H)$ formalism can be used to describe both linear and nonlinear input-output systems. Working in the usual Markov limit of quantum optics as well as an instantaneous coupling limit for the waveguide interconnections~\cite{GJ09b,Goug10}, a Gough-James circuit model can be collapsed via operator-algebraic manipulations into a single $(S,L,H)$ triple for the entire circuit. The Master Equation, Stochastic Schr\"{o}dinger Equation, or Heisenberg (Hudson-Parthasarathy) equations of motion~\cite{GZ04,WM10} for both internal observables and output fields can be derived from the circuit $(S,L,H)$ parameters for analysis and (for systems of modest size) numerical simulation~\cite{Teza12,GJ09a}.

In traditional electrical engineering the port connection topology of a collection of components is usually called a netlist. Several widely-used conventions exist for text-based specification of the netlist of a circuit according to some formal grammar. For example, VHSIC Hardware Description Language (VHDL)~\cite{Pedroni} and Verilog~\cite{Thomas} are formats intended mainly for use in digital electronics, while Modelica~\cite{Frit2004} is a more recent format designed to accommodate multiple physical domains. Any of these text-based netlist formats can be used to specify the port connection topology of a photonic, optomechanical or optoelectronic circuit, with the practical advantage that such formats can be generated and read by graphical-user-interface circuit design software such as {\tt gschem} or OpenModelica. It is natural to consider the task of computer-automated `parsing' of a text-based netlist specification to produce a Gough-James circuit expression, which can then be reduced algebraically to a time-domain model for the quantum stochastic circuit dynamics. We have previously demonstrated~\cite{Teza12} such a schematic capture workflow, based on VHDL, for the construction of quantum circuit models. In this article we move on to consider the {\em transformation} of quantum photonic circuit models as a fundamental methodology for verification and robustness analysis, working with Modelica rather than VHDL as a netlist specification format because of its simplified grammar and its integration with \emph{Mathematica}~\cite{Wolfram}, which we will use as a computational engine for symbolic manipulation.

\begin{figure}[h]
\begin{center}\includegraphics[width=0.85\textwidth]{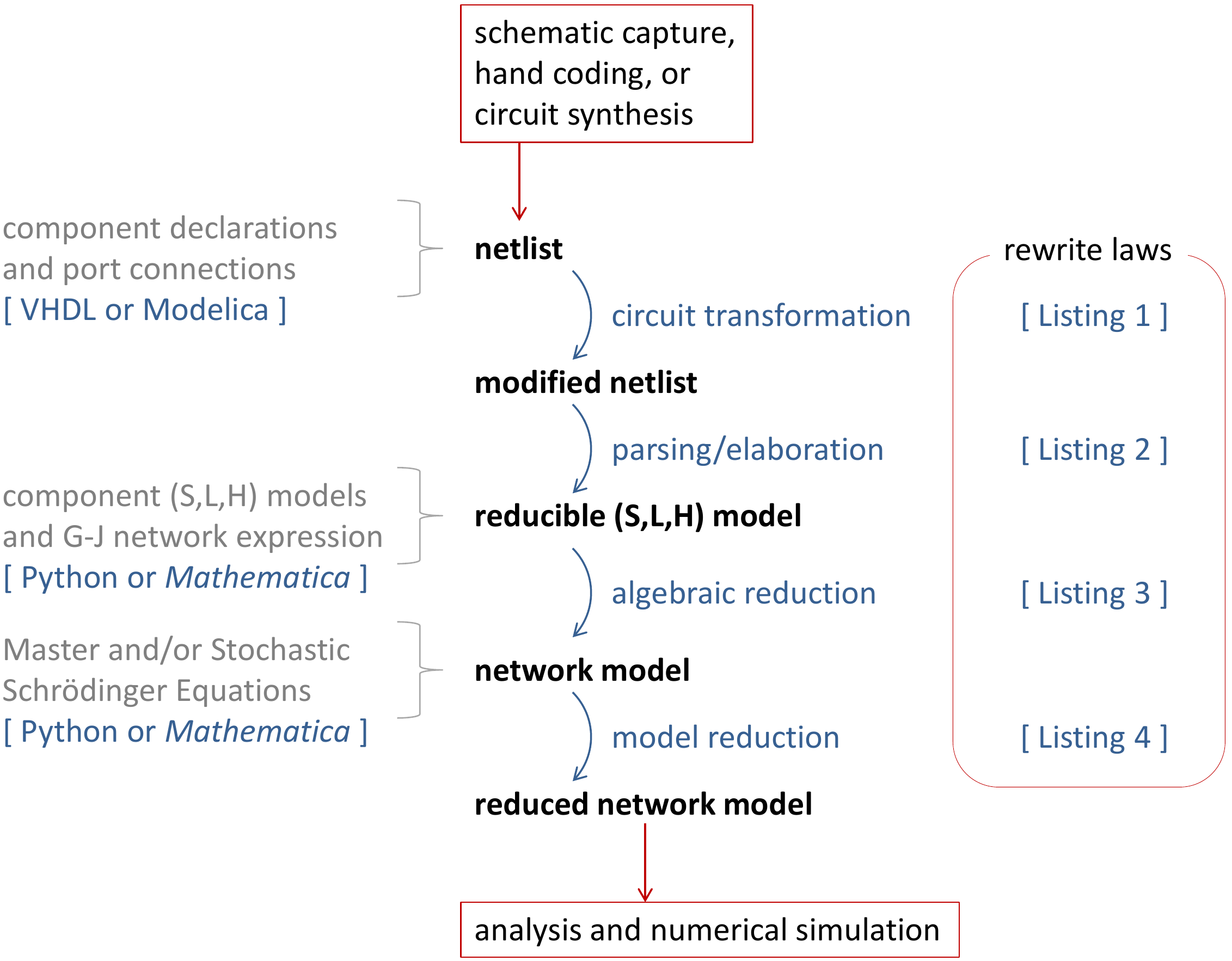}\end{center}
\caption{Quantum photonic circuit analysis workflow, viewed as a series of transformations of a circuit model. While some rewrites are most naturally applied at the level of the netlist term algebra, others must be done in the algebra of Hilbert-space operators. The Gough-James algebra provides a natural intermediate representation of the circuit model. Listings 1-4 can be found in the Supplementary Data.}\label{fig:flowchart}
\end{figure}

An outline of our current circuit analysis approach is presented in Fig.~\ref{fig:flowchart}. An initial netlist description of the circuit is produced using schematic capture or coded by hand; an important goal for future work in this field will be the algorithmic synthesis of circuit topologies implementing a desired function. Some circuit transformations will most naturally be applied at the netlist level of description---below we will consider the addition of optical propagation losses as an example. The initial or modified netlist can then be rewritten into the Gough-James algebra of series and concatenation products of $(S,L,H)$ component models, and the Gough-James circuit model can in turn be reduced to an overall $(S,L,H)$ model for the entire circuit. Further transformations of the circuit model, such as adiabatic elimination of fast dynamics, are then implemented via manipulations of the Hilbert-space operators appearing in the circuit scattering matrix, coupling vector and Hamiltonian. Final equations of motion can then be extracted for symbolic analysis and/or numerical simulation.

It is interesting to note that all the above circuit model transformations can be regarded as applications of compact sets of rewrite laws in a term rewriting system (TRS)~\cite{TeReSe}. As noted above, we have chosen to implement such rewrite laws within \emph{Mathematica}. The study of TRSs provides a common framework for abstract algebra, the theory of computation, and formal verification methods and is an active area of research in contemporary engineering. As it appears that photonic circuit models of the type we consider here can be treated on equal footing, it seems reasonable to hope that sophisticated TRS-based tools being developed for classical synthesis and verification~\cite{Arvi99} could be adapted for use in quantum engineering as well.

In recent work~\cite{Kerc10,Kerc11} we have proposed a class of quantum photonic circuits that autonomously implement a form of quantum error correction (QEC) based on stabilizer coding, continuous syndrome measurement and restorative feedback. No external control or clocking signals are required; once fabricated according to specification the circuit should continuously implement the QEC protocol by virtue of the fixed Hamiltonian couplings among its cavity QED-based components and optical waveguides. The circuit is powered by coherent laser inputs, whose frequencies should be accurately stabilized but whose amplitudes need only respect a certain parameter hierarchy. Such QEC models present a useful set of elementary examples for quantum circuit theory incorporating features such as coherent optical signals, component-component entanglement and feedback loops.

An elementary question to ask about these quantum memory circuits is how their performance would be degraded by propagation losses in the waveguides. While it is straightforward in principle to perform this type of robustness analysis via numerical simulation of modified quantum circuit models, the complexity of even the ideal (lossless) models is such that adding loss terms by hand would be prohibitively tedious. In order to obtain a valid quantum optical model for the circuit with propagation losses, each port-to-port connection in the lossless model should be replaced by a compound connection in which the original upstream port is connected to one input port of a beam-splitter and the corresponding beam-splitter output port is then connected to the original downstream port. The reflection coefficient of the beam-splitter sets the effective propagation loss of the connection, and each such addition of an unconnected output port (corresponding to the beam-splitter reflection) to the circuit model increases the number of Lindblad terms in the overall Master Equation.

In order to illustrate how this type of transformation can be performed automatically in our circuit analysis workflow, we first display a segment of Modelica code specifying the netlist for a simple sub-circuit that implements a continuous two-qubit parity measurement~\cite{Kerc09}:

\begin{lstlisting}[label={lst:Circuit}]
model TwoQubitParity
  Photonics.Components.CoherentField W(Amplitude=alpha);
  Photonics.Components.SingleCavity Q1(CavityType=Zprobe, HilbertSpace=Q1);
  Photonics.Components.SingleCavity Q2(CavityType=Zprobe, HilbertSpace=Q2);
equation
  connect(W.output1,Q1.input1);
  connect(Q1.output1,Q2.input1);
end TwoQubitParity;
\end{lstlisting}

\noindent The code specifying the {\footnotesize\tt TwoQubitParity} sub-circuit begins with a set of declarations (the three lines prior to the {\color{Green}\footnotesize\bf\texttt{equation}} keyword) of the three components it contains: a coherent field input {\footnotesize\tt W} and qubit-cavity components {\footnotesize\tt Q1} and {\footnotesize\tt Q2}. The two lines after the {\color{Green}\footnotesize\bf\texttt{equation}} keyword specify the architecture via simple statements of which output ports are connected to which input ports. In order to insert a propagation loss between {\footnotesize\tt Q1} and {\footnotesize\tt Q2}, it suffices simply to rewrite the netlist specification by adding the line {\footnotesize\tt Photonics.Components.Loss L(LossParam=theta);} to the declaration block and substituting the line {\footnotesize\tt connect(Q1.output,Q2.input1);} in the architecture block with the lines {\footnotesize\tt connect(Q1.output1,L.input1);} and {\footnotesize\tt connect(L.output1,Q2.input1);}. Clearly, such manipulations of the netlist specification code can be implemented straightforwardly using pattern matching and string replacement. We provide example \emph{Mathematica} code for this purpose in Listing 1 of the Supplementary Data for this article.

\begin{figure}
\begin{center}\includegraphics[width=.75\textwidth]{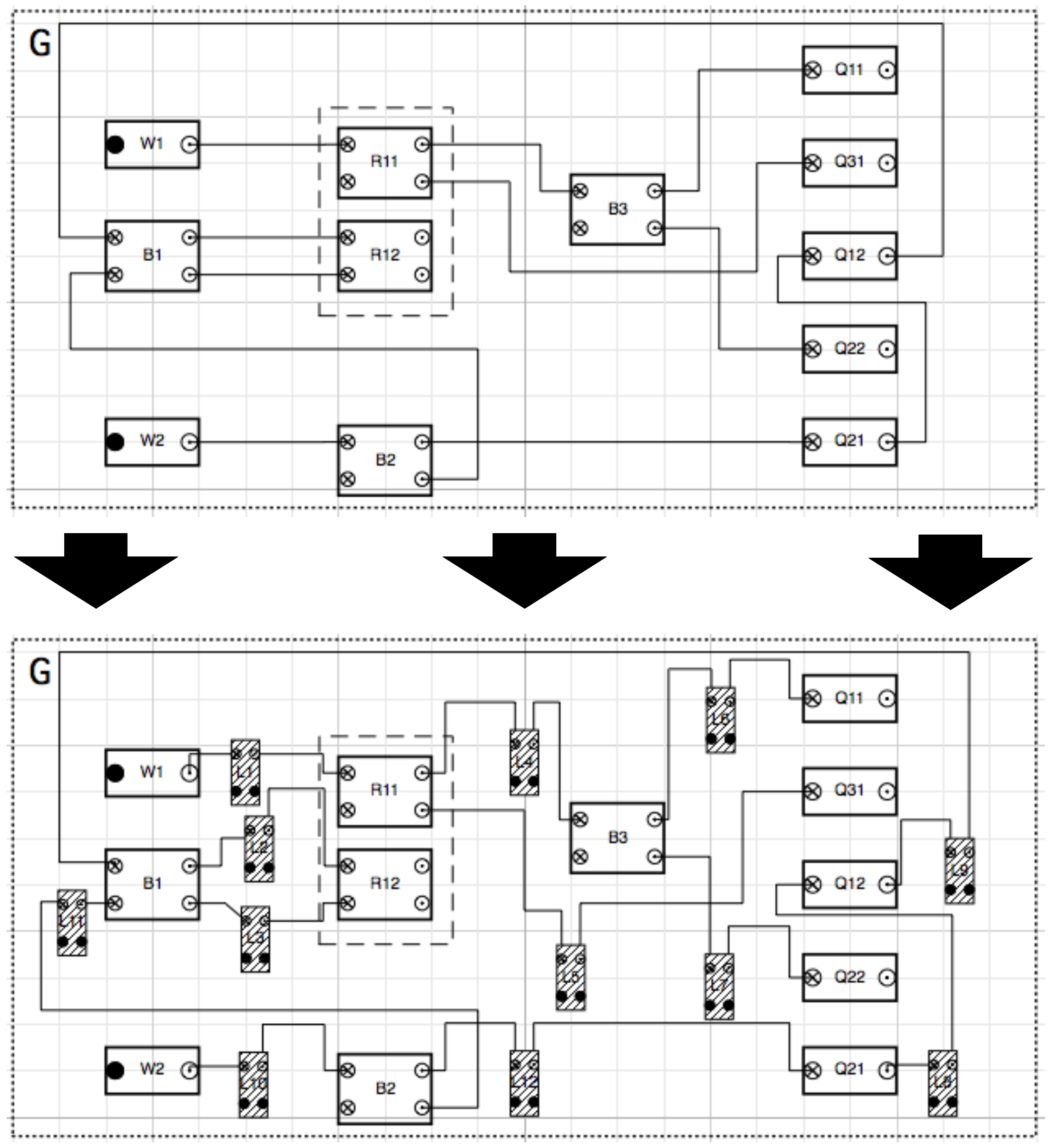}\end{center}
\caption{\label{fig:bitloss} TOP: Screen capture from the Modelica system designer of a graphical representation of half the photonic circuit for a quantum memory based on the bit-flip/phase-flip code. This diagram is analogous to the schematic in Appendix C of~\cite{Kerc10}. BOTTOM: Corresponding circuit representation including propagation losses.}
\end{figure}

A larger-scale example of the propagation loss transformation is depicted in Fig.~\ref{fig:bitloss}. The upper panel presents a screen capture (from the Modelica system designer) of a graphical representation of half the photonic circuit for an autonomous quantum memory based on the bit-flip/phase-flip code, without propagation losses. The lower panel shows the same sub-circuit after a transformation inserting beam-splitters into every port connection to enable rigorous modeling of propagation losses. We would like to emphasize that although this type of loss-insertion transformation is quite simply accomplished via rewriting of the netlist specification, it would be far more complicated to implement at the subsequent levels of model representation as a Gough-James circuit expression or overall $(S,L,H)$ triple.

Proceeding to the next stage of the circuit analysis workflow, we utilize a \emph{Mathematica} script to rewrite the final connectivity model from the netlist term algebra to the Gough-James algebra. Terms in the Gough-James algebra are constructed from constants representing the components, and operators indicating the connections among components. Infix notation is used with parentheses for clarity. The elementary operators for connecting components are the series product (denoted $\triangleleft$), with $B\triangleleft A$ indicating that all output ports of component $A$ are connected to corresponding input ports of component $B$, and the concatenation product (denoted $\boxplus$), with $D\boxplus C$ indicating that components $D$ and $C$ coexist in the circuit but have no connections. The result of a series or concatenation product can be treated as a new component. The input/output ports of $B\triangleleft A$ are the input ports of $A$ and the output ports of $B$; $D\boxplus C$ has the input and output ports of both $C$ and $D$. In practice it is useful to add a permutation (crossover) operator that reorders the output ports of a component, as well as a feedback operator $[M]_{i\rightarrow j}$ that connects output port $i$ of a component $M$ to its own input port $j$. It is generally also necessary to utilize ancillary $n$-line `pass-through' components $I_n$ to construct a complete circuit expression. For example, if $A$ has two output ports and $B$ has four input ports, a connection of the outputs of $A$ to the first two inputs of $B$ without any other connections would be written $B\triangleleft(A\boxplus I_2)$.

The minimal task in this stage of the analysis is thus to replace the list of port-to-port connections in the Modelica architecture block with component-to-component connections, inserting ancillary pass-through or permutation blocks as necessary. This is not a one-to-one mapping---many distinct Gough-James circuit expressions can faithfully represent a given netlist. All such expressions are equivalent in that the application of algebraic reduction rules (see below) will bring any such equivalent circuit expression to a unique normal form, the overall $(S,L,H)$ triple for the circuit. It is useful however to consider strategies for obtaining relatively compact Gough-James circuit expressions, for ease of inspection and also to minimize the complexity of the subsequent algebraic reduction. The algorithm we use is based on the idea of trying to group together components that are the most connected. For example, if two 2-port components are connected by {\footnotesize\tt connect(A.output1,B.input1);} and {\footnotesize\tt connect(A.output2,B.input2);} we can bring them cleanly into the Gough-James circuit expression as $(B \triangleleft A)$. The algorithm assigns a score to each pair of connected components according to how fully connected the two circuit elements are. The higher the score, the more certain the algorithm is that grouping the two elements together will lead to a compact circuit expression. The parsing algorithm then finds the highest-scoring connector joining the two elements, replacing them with either a series or feedback product (padding with pass-through components as necessary), and repeating the process over and over until we have accounted for all of the netlist connections in a single Gough-James expression (using $\boxplus$ to join together any disconnected sub-nets).

\begin{figure}[h]
\centering
\includegraphics[width=1.00\textwidth]{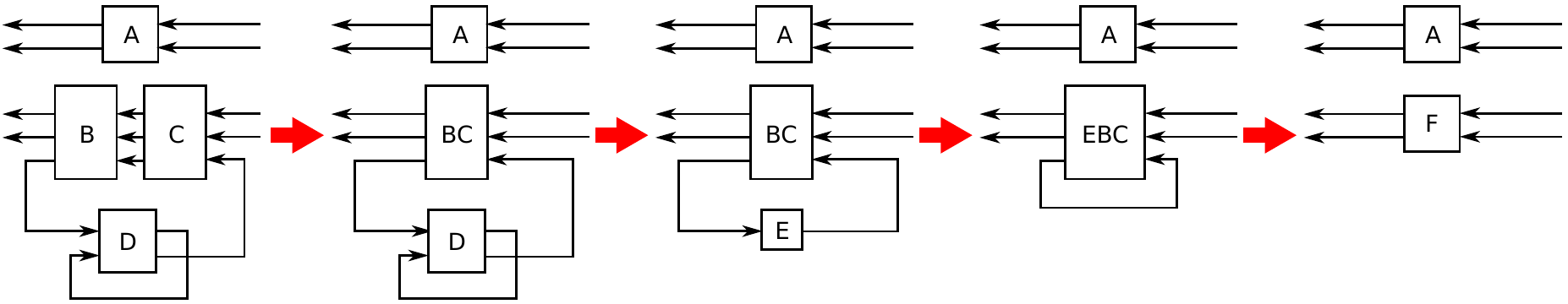}
\caption{Steps the circuit parser takes to convert a simple netlist into a Gough-James circuit expression.}
\label{fig:parser}
\end{figure}

As an example, consider the circuit diagram at the left of Fig.~\ref{fig:parser}. In the first step, the parser replaces $B$ and $C$ with $BC = B \triangleleft C$ (the pair has a `score' of 1.0 because all of the $C$ outputs match inputs in $B$). The remaining steps are analogous:

\begin{table}[h]
\centering
\begin{tabular}{c|c|c}
	Step & Replacement & Score \\ \hline
	1 & $BC = B \triangleleft C$ & $1.00$ \\
	2 & $E = [D]_{1\rightarrow 2}$ & $1.00$ \\
	3 & $EBC = (I_2 \boxplus E) \triangleleft BC$ & $0.33$ \\
	4 & $F = [EBC]_{3\rightarrow 3}$ & $1.00$
\end{tabular}
\end{table}

\noindent The final circuit expression, $A \boxplus F$, can be read off from the substitutions in the table above. It is:
\begin{equation}
	\label{eq:parserout}
	A \boxplus \bigl[\left(I_2 \boxplus [D]_{1\rightarrow 2}\right) \triangleleft B \triangleleft C\bigr]_{3\rightarrow 3}
\end{equation}
\emph{Mathematica} code for implementing this algorithm is included in Listing 2 of the Supplementary Data.

Algebraic reduction of a Gough-James circuit expression is performed by applying normal quantum-mechanical operator algebra plus the following rewrite rules~\cite{GJ09a}:
\begin{eqnarray}
B \triangleleft A &\rightarrow& \left( S_BS_A,L_B + S_BL_A,H_B + H_A + {\rm Im}\{L_B^\dag S_BL_A\}\right),\\
B \boxplus A &\rightarrow& \left( S_B\oplus S_A,L_B\oplus L_A,H_B+H_A \right).
\end{eqnarray}
Here $\oplus$ denotes the usual direct sum of matrices or vectors. An analogous rule for the feedback operation $[A]_{i\rightarrow j}$ is described in~\cite{Teza12}. Here $(S_B,L_B,H_B)$ is the parameter triple for component $B$ and $(S_A,L_A,H_A)$ is the triple for component $A$. We assume that the software can obtain component parameter triples in symbolic form from a library. We include a \emph{Mathematica} script that implements the above rewrite rules in Listing 3 of the Supplementary Data. When these rules have been applied to completion, the original Gough-James expression is replaced by a single $(S,L,H)$ triplet that represents the entire circuit. In general the circuit $(S,L,H)$ expression can be rather unwieldy and may not be amenable to intuitive interpretation---its parameters summarize the coupled quantum dynamics of all the components in the circuit in a way that makes it straightforward to extract overall evolution equations for numerical simulation, but analytic verification of the circuit behavior will generally require further model reduction steps.

\begin{figure}[tbh]
\begin{center}\includegraphics[width=\textwidth]{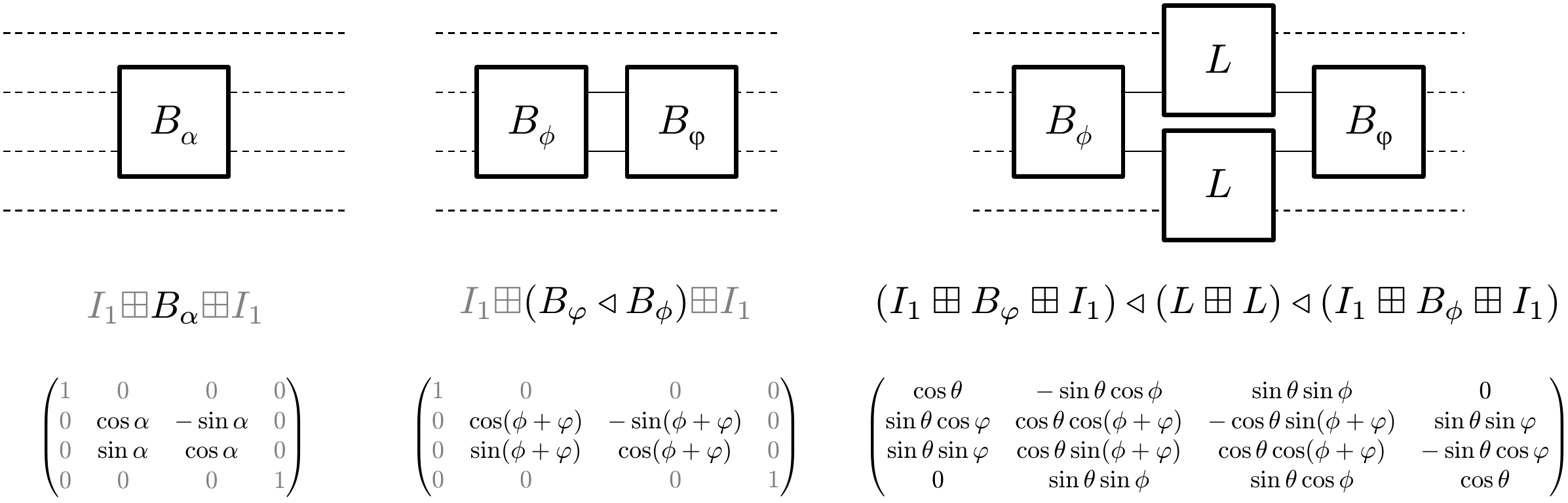}\end{center}
\caption{Example of equivalent circuit $(S,L,H)$ models (left and center columns) corresponding to distinct netlists that transform differently under addition of propagation losses. The top row shows circuit diagrams, the middle row shows the corresponding Gough-James expressions, and the bottom row shows the overall scattering matrices $S$ for each circuit.}\label{fig:reducts}
\end{figure}

We are now in a position to elaborate on an earlier comment regarding the advantage of considering circuit model transformations very early in the analysis workflow. The top-left and top-center diagrams in Fig.~\ref{fig:reducts} depict a simple beamsplitter and a compound beamsplitter that can be formed by a Mach-Zehnder type connection topology~\cite{Teza12}. The corresponding Gough-James expressions and scattering matrices (which are the only non-zero components of the $(S,L,H)$ triples for such simple circuits) are shown in the middle and bottom rows of Fig.~\ref{fig:reducts}. If $\phi+\varphi=\alpha$ the simple and compound beam-splitters are equivalent photonic circuits (in the instantaneous-coupling limit, without propagation losses). The single beamsplitter circuit does not change under a transformation rule that adds losses to internal port-to-port connections only, while the compound beamsplitter circuit is transformed to the circuit described in the right column of Fig.~\ref{fig:reducts}. Here $\theta$ is a loss parameter and $L=B_\theta$; note that we recover the lossless scattering matrix as $\theta\rightarrow 0$. This example clearly illustrates that some information about internal port-to-port connections is lost by the time the circuit model has been reduced to an overall $(S,L,H)$ parameter triple, implying that some important types of circuit transformations (such as the addition of propagation losses) cannot be implemented directly on the final $(S,L,H)$ model but rather must be implemented at an earlier stage of the rewrite chain.

Once an overall circuit $(S,L,H)$ model has been obtained, a final class of transformations (which could not have been performed at the netlist or Gough-James levels of representation) may be applied via rewrites in the operator algebra. For example, our analyses of the quantum memory circuits proposed in~\cite{Kerc10,Kerc11} have relied on a limit theorem for QSDEs~\cite{Goug10,Bout08} to produce reduced models for the overall network that are amenable to behavior verification and tractable for numerical simulation. Practically, application of the limit theorem requires that certain operator products be computed which correspond to the coefficients of a limit QSDE for the slow degrees of freedom in an open quantum system (see Section 2.2 of~\cite{Bout08} for general results and~\cite{Kerc10} for specific application to our QEC circuit models). Once the limiting subspace has been defined and the corresponding structural requirements have been verified, computation of the limit QSDE requires a straightforward but potentially cumbersome series of algebraic manipulations that can easily be automated using pattern matching and string replacement; a sample \emph{Mathematica} script for this purpose is provided in Listing 4 of the Supplementary Data.

\begin{figure}[tbh]
\begin{center}\includegraphics[width=0.9\textwidth]{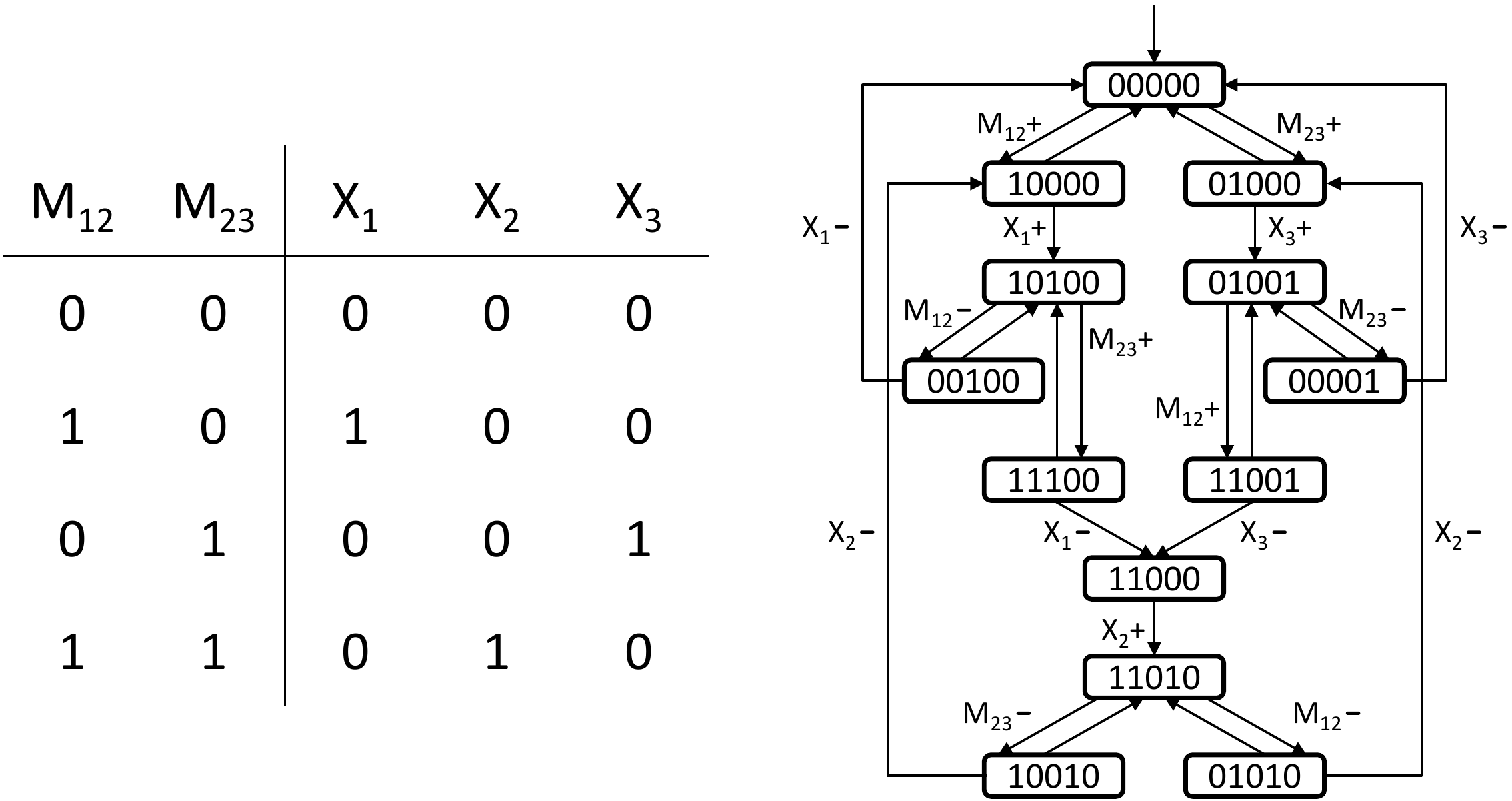}\end{center}
\caption{LEFT: Table of possible input/syndrome signal values $(M_{12},M_{23})$ and desired output/correction $(X_1,X_2,X_3)$ signal values for the desired bit-flip QEC controller~\cite{Kerc10}. RIGHT: Asynchronous Transition System~\cite{Yako92,Cort97} specifying the desired controller behavior.}\label{fig:states}
\end{figure}

The left table of Fig.~\ref{fig:states} shows the well-known control scheme for the bit-flip QEC protocol. Here $M_{ij}$ is a measurement signal taking values in $\{0,1\}$ that indicates the presence of an odd parity condition between qubits $i$ and $j$ in the quantum register. For the three-qubit bit-flip code, two such syndrome measurements are sufficient to localize an error. Whereas $M_{12}$ and $M_{23}$ are input signals to the controller, we use $X_1$, $X_2$ and $X_3$ to indicate the values of controller output signals that drive corrective bit-flip actions on qubits $1$, $2$ and $3$, respectively. Each row of the table indicates the required configuration of the controller output signals $X_k$ when the input signals are as indicated in the first two columns. The diagram on the right side of Fig.~\ref{fig:states} presents a Transition System (TS)~\cite{Yako92,Cort97} as a full (asynchronous) specification of the desired behavior of the controller. In the TS diagram, which has the form of a graph with labeled directed edges, each node represents a state of the controller, the directed edges indicate possible transitions between states, and the label on any given edge indicates a corresponding signal transition. The binary string specifying each controller state corresponds to the values of the signals $M_{12}$, $M_{23}$, $X_1$, $X_2$ and $X_3$ in order. The state and transition labels are thus redundant and the latter have been omitted in a few places to avoid cluttering the diagram.

After transformation via the QSDE limit theorem mentioned above, the $(S,L,H)$ model (without propagation losses) for our bit-flip QEC circuit~\cite{Kerc10} contains the following Lindblad operators and Hamiltonian terms pertaining to the behavior of the controller:
\begin{eqnarray}
L_{s1} &=& \alpha\left( \sigma_+^{R1}M_{12}-\Pi_0^{R1}(1-M_{12})\right),\\
L_{r1} &=& \alpha\left( -\Pi_1^{R1}M_{12}+\sigma_-^{R1}(1-M_{12})\right),\\
L_{s2} &=& \alpha\left( \sigma_+^{R2}M_{23}-\Pi_0^{R2}(1-M_{23})\right),\\
L_{r2} &=& \alpha\left( -\Pi_1^{R2}M_{23}+\sigma_-^{R2}(1-M_{23})\right),\\
H_c &=& \Omega\left( \sqrt{2}X^{Q1}\Pi_1^{R1}\Pi_0^{R2} + X^{Q2}\Pi_1^{R1}\Pi_1^{R2} - \sqrt{2}X^{Q3}\Pi_0^{R1}\Pi_1^{R2}\right).
\end{eqnarray}
Here $\vert\alpha\vert^2$ represents the strength (photons per unit time) of a probe optical field used to monitor the syndromes of the quantum register and $\Omega$ is a parameter for the feedback strength~\cite{Kerc10}. Our QEC controller circuit utilizes a pair of set-reset relay components~\cite{Mabu09} $R_1$ and $R_2$ driven by the syndrome inputs $M_{12}$ and $M_{23}$ to switch the output signals $X_1$, $X_2$ and $X_3$. In the above expressions $\Pi_m^{Rn}$ denotes a projection operator into state $m$ for $R_n$, $\sigma_{+(-)}^{Rn}$ is a raising (lowering) operator for $R_n$, and $X^{Qn}$ is a Pauli $\sigma_x$ operator for register qubit $n$. The Lindblad terms above implement the responses of the relay states to the syndrome inputs. For example, in a Master Equation or Stochastic Schr\"{o}dinger Equation for the QEC circuit, $L_{s1}$ and $L_{r1}$ will contribute dynamical terms that cause $R_1$ to decay exponentially (with rate $\vert\alpha\vert^2$) to state $0$ when $M_{12}=0$ and to state $1$ when $M_{12}=1$. The remaining Lindblad terms will do the same for $R_2$ and $M_{23}$. The three terms in the control Hamiltonian $H_c$ implement corrective feedback on the register qubits whenever the states of the relays are not both $0$, following precisely the scheme given in the table of Fig.~\ref{fig:states}. We thus see that the behavior of our bit-flip QEC circuit can be verified by inspection relative to a conventional asynchronous controller specification such as the TS diagram of Fig.~\ref{fig:states}. We wish to emphasize, however, that this type of transparent correspondence between the $(S,L,H)$ terms and desired TS behavior emerges only after the QSDE limit transformation in the analysis of our QEC circuits.

\begin{figure}[tbh]
\begin{center}\includegraphics[width=.75\textwidth]{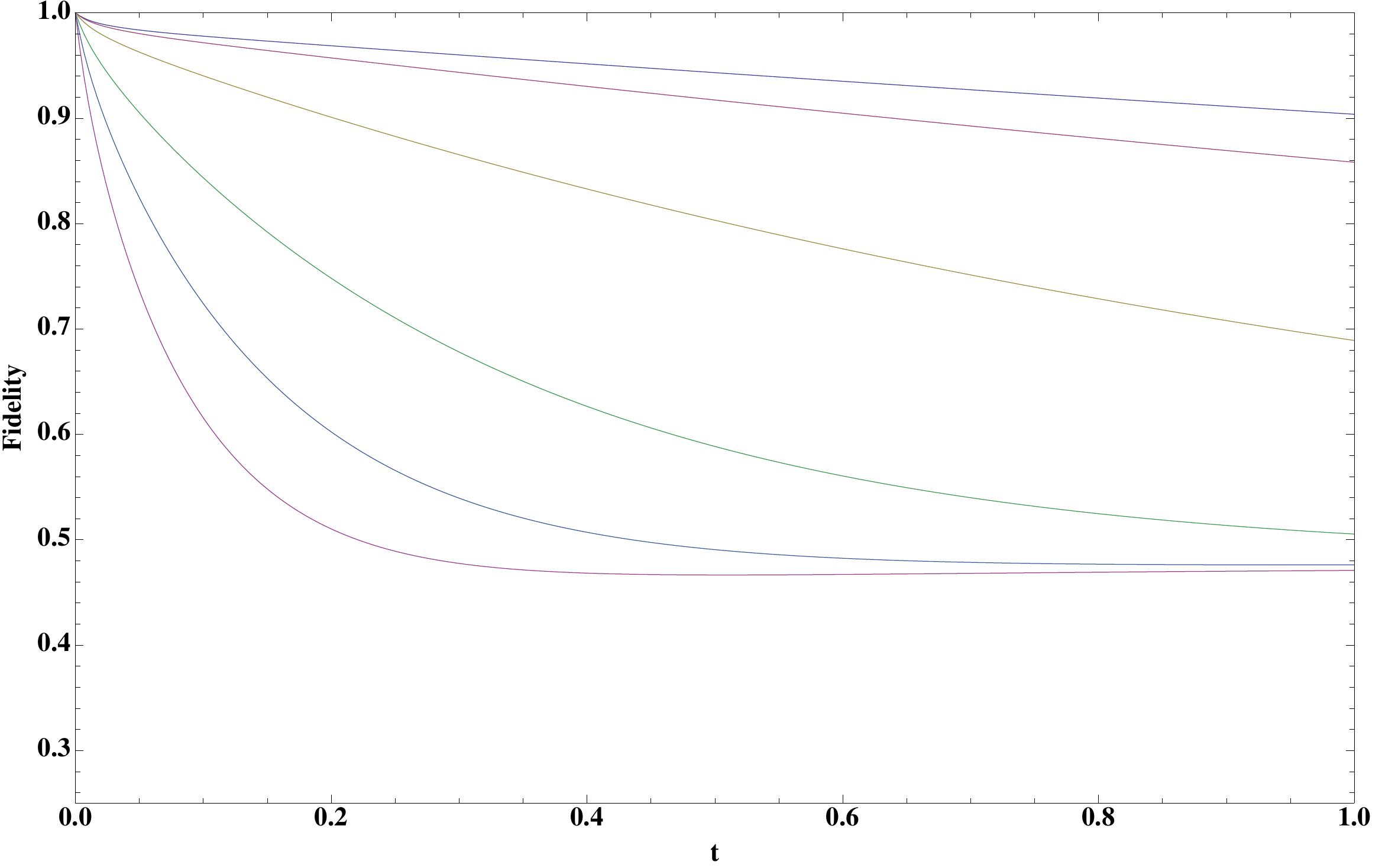}\end{center}
\caption{\label{fig:lossy-bit-flip} Decay of fidelity $\langle \Psi_{0} | \rho_{t} | \Psi_{0}\rangle$ for 3-qubit bit-flip code with loss parameters $\theta=\{0,1,2.5,5,7.5,10\}\pi/1000$ (top to bottom curves) for a bit-flip QEC circuit.  For consistency with \cite{Kerc10}, the feedback strength $\Omega = \frac{|\beta|^{2} \gamma}{2\Delta}$ is set to a constant value of $210$.}
\end{figure}
\begin{figure}
\begin{center}\includegraphics[width=.75\textwidth]{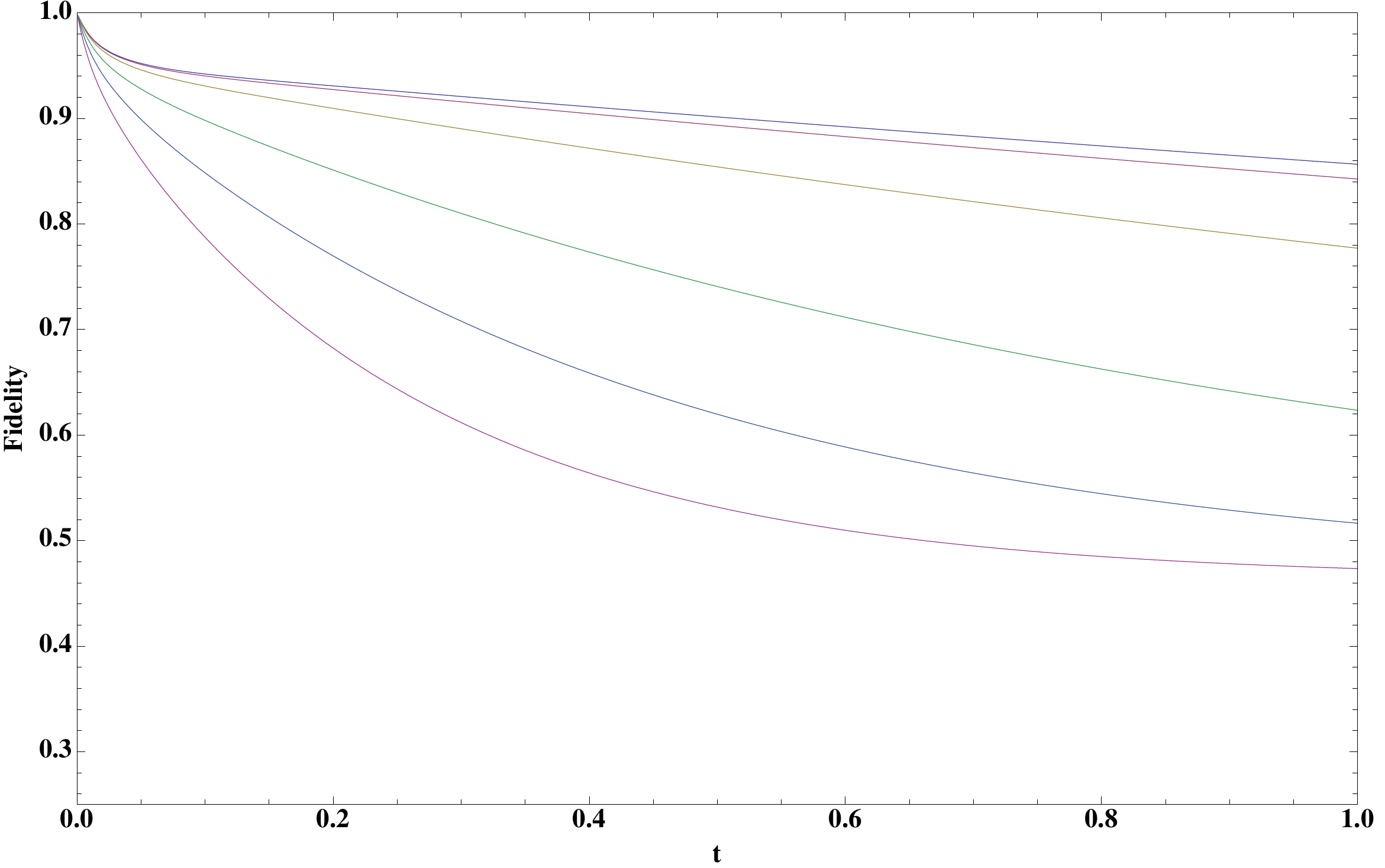}\end{center}
\caption{\label{fig:lossy-bacon-shor} Decay of fidelity $\langle \Psi_{0} | \rho_{t} | \Psi_{0}\rangle$ for 9-qubit Bacon-Shor code with loss parameters $\theta=\{0,1,2.5,5,7.5,10\}\pi/1000$ (top to bottom curves) for a nine-qubit QEC circuit. For consistency with \cite{Kerc11}, the feedback strength $\Omega = \frac{|\beta|^{2} \gamma}{2\Delta}$ is set to a constant value of $200$ for each of these runs.}
\end{figure}


If the analysis workflow is repeated for our QEC circuit models with a propagation-loss transformation inserted at the netlist level, new Lindblad operators are generated such as
\begin{equation}
L_{pl} = \alpha\theta Z^{Q2},\label{eq:lbit}
\end{equation}
for the bit-flip case, and
\begin{equation}
L_{pl} = \alpha\theta Z^{Q2}Z^{Q5}Z^{Q8},\label{eq:lnine}
\end{equation}
for the nine-qubit code. Here $\theta$ is a propagation loss parameter and $Z^{Qj}$ is the Pauli $\sigma_z$ operator for qubit $j$ in the quantum register. In both cases we see that nonzero optical propagation loss causes the appearance of new error processes that are not corrected by the original QEC code---a phase error in the bit-flip circuit, and a three-qubit correlated phase error in the nine-qubit circuit. This is of course not a pathology of our photonic QEC implementation, but rather an intrinsic property of the codes used---analogous error terms arise in a conventional approach as a consequence of ancilla decoherence while syndrome observables are being accumulated by a sequence of two-qubit gates. Additional loss-induced modifications of the QEC circuit dynamics are computed automatically by the model rewriting workflow.

To assess the quantitative impact of propagation losses we can numerically integrate the Master Equation corresponding to the (reduced) circuit $(S,L,H)$~\cite{Teza12}. As shown in Fig.~\ref{fig:lossy-bit-flip} and Fig.~\ref{fig:lossy-bacon-shor}, the fidelity of the encoded qubit decays more rapidly as the propagation loss parameter $\theta$ is increased. For simplicity, we here have assigned the same loss parameter to each port-to-port connection in the initial netlist. We wish to emphasize that this level of quantitative analysis for the nine-qubit code would be practically intractable without the automated circuit analysis workflow that we have outlined in this paper. An excerpt of the netlist and corresponding Gough-James circuit expression are included in the Supplementary Data for this article, together with the coupling vector and Hamiltonian operator from the overall circuit $(S,L,H)$ for the reduced model.

In conclusion, we have described a model transformation workflow for analyzing complex quantum photonic circuits and have illustrated key concepts using examples related to prior work on autonomous quantum error correction. Code listings are provided in the Supplementary Data for this article to demonstrate how the model transformations can be implemented via compact sets of rewrite rules. A practical approach to analyzing the functional robustness of a photonic circuit to propagation losses in its internal waveguide connections has been presented with numerical results for bit-flip and nine-qubit QEC models. And finally, we have introduced the possibility of formal verification via $(S,L,H)$ analysis of quantum photonic circuit behavior relative to conventional (in contemporary electrical engineering) specification formats for asynchronous controllers.

\ack
This work has been supported by the NSF (PHY-1005386), AFOSR (FA9550-11-1-0238) and DARPA-MTO (N66001-11-1-4106). R.~Hamerly is supported by the NSF GRFP and a Stanford Graduate Fellowship. D.~S.~Pavlichin is supported by a Stanford Graduate Fellowship.

\end{document}